# Is the Indian Stock Market efficient – A comprehensive study of Bombay Stock Exchange Indices

Achal Awasthi[1], Dr. Oleg Malafeyev [2]

1 - Undergraduate Student, Department Of Physics, Shiv Nadar University, India. Postal Address – C-23 Dhawalgiri Apartments Sector 11, Noida, India.
Tel – (+91)9999493777, E-Mail: aa777@snu.edu.in

2 - Professor and Chair (Socioeconomic systems Modeling Department), Faculty of Applied Mathematics and Control Processes, Saint Petersburg State University, Russia
Tel - (812) 428-42-47, E-mail: malafeyevoa@mail.ru

**Abstract –**

How an investor invests in the market is largely influenced by the market efficiency because if a market is efficient, it is extremely difficult to make excessive returns because in an efficient market there will be no undervalued securities. However, there is a possibility of making excess returns if the market is not efficient. This article analyses the five popular stock indices of BSE. This would not only test the efficiency of the Indian Stock Market but also test the random walk nature of the stock market. The study which was undertaken in this paper has provided strong evidence in favour of the inefficient form of the Indian Stock Market. This article also covers the post-election period when the new government came into power and boosted the investors' confidence. The series of stock indices in the Indian Stock Market are found to be biased random time series and the random walk model can't be applied in the Indian Stock Market. This study confirms that there is a drift in market efficiency and investors can capitalize on this by correctly choosing securities that are undervalued.

## 1. Introduction –

Mathematical modeling approach is widely used in different areas of science and technology (see for example [1-3]).It provides deep insights if try to mathematically model various phenomenon related to economics and finance (see for example [4-17]). The basics of the approaches and methodologies applied here can also be seen in [18-30].

An efficient market can be described as a market, where the market price represents an unbiased estimate of the actual value of the investment. It is not necessary that the market price and the actual value must be same at each point of time. The essential condition is that the errors in the market price should be unbiased i.e. the market prices can be different from the actual value but if these deviations are random, the market would be efficient. It strongly influences the investment strategy of an investor because it's extremely difficult to choose undervalued securities in an efficient market as there are no undervalued securities in an efficient market. However, in an inefficient market, an investor can make excessive returns. In this paper, five important stock indices are analyzed by using non-parametric tests. This would not only test the efficiency of the stock market but also test the random walk nature of the stock market. The main contribution of this paper is to comprehensively test the random walk theory of stock market prices, by testing the monthly returns of the five popular stock indices of BSE for the null hypothesis of a random walk. It adds on previous studies for the Indian stock market. This paper is important because it covers the period before the 2014 General Elections and also the period after that. In the 2014, General Elections, the Bhartiya Janta Party beat the ruling Congress Party by a huge amount and there was a shift in the governmental regime. This shift was also seen in the Indian stock market as now the investors were more confident which boosted the stock market.

## 2. Literature Survey –

There are some previously conducted studies related to the analysis of stock indices which are summarized below.

Anand Pandey (2003) tested the efficiency of the three popular stock Indices of National Stock Exchange using non parametric tests. It was found that the time series corresponding to stock indices in the Indian Stock Market was biased random time series.

V. N. Kolokoltsov, O.A. Malafeyev in their books Understanding Game Theory (2010) and The mathematical Analysis of the Financial Markets (2010) analyzed the mathematical aspects of financial markets. The stability equations have been useful in many fields of studies including physics, finance, bio-science and economics. Zubov (2015) in his paper, Stability of quasilinear dynamic systems with after effect examines the robust stability of control systems and also discusses the issues of existence and stability of stationary modes in dynamic systems with aftereffect.

Maria Rosa Borges (2011) tested the efficiency of the Lisbon stock market and concluded that the Portuguese stock market index has been approaching random walk behavior since 2000.

This study was carried out so that the small and medium investors could be motivated to save and invest in the capital market only if their securities in the market are appropriately priced. But many investors are unaware of the fact that not all indices in the Indian Stock Market are worth investing in. The previous studies have tested the efficiency in the global stock markets and also tested the random walk for various popular indices. But in India, very few studies have examined the weekly and monthly returns of the stock market in particular stock indices, like CNX Defty, BSE 200 Index, and Nifty Junior.

Thus, in the Indian context, apart from a few studies, the available evidence in general points out to the fact that the consecutive price changes are independent and the trends in movement of stock prices in Indian markets can be explained by random walk model. In this paper we are particularly interested in the pre and post-election period of 2014, both because it has not been covered by previous studies, and it is being covered in this paper.

## 3. Methodology –

Since the test of weak forms of EMH, in general, have come from the random walk literature, so it will be tested whether or not successive price changes were independent of each other. In this paper, Autocorrelation and Runs Test will be used to test the efficiency of the stock market.

### 3.1 Autocorelation ACF (k)

The autocorrelation function ACF (k) for the time series $Y_t$ and the k-lagged series $Y_{t-k}$ is defined as:-

$$\text{ACF (k)} = \frac{\sum_{(t=1-k)}^{n}(y_t - \bar{y})(y_{t-k} - \bar{y})}{\sum_{(t=1)}^{n}(y_t - \bar{y})^2} \qquad (1)$$

where $\bar{y}$ is the overall mean of the concerned series with 'n' observations.

The standard error of ACF (k) is given by:-

$$se_{ACF(k)} = \frac{1}{\sqrt{n-k}} \qquad (2)$$

If n is sufficiently large (n ≥50), the standard error of ACF (k) can be approximated to:-

$$se_{ACF(k)} = \frac{1}{\sqrt{n}} \qquad (3)$$

The following 't' distribution will be used to test the hypothesis whether ACF (k) is significantly different from zero or not.

$$t = \frac{ACF(k)}{se_{ACF(k)}} \qquad (4)$$

A random walk series drifts up and down over time. In some situations it may be difficult to judge whether a trend or drift is occurring. Thus, in order to determine whether a series has significant trend or whether it is a random walk, the t-test is applied on the series of first differences.

### 3.2 Runs Test –

Runs Test is a non-parametric test and depends only on the sign of the price changes but not on the magnitude of the price. It will be used to decide if a data set is from a random process or not. It depends only on the sign of the price. The test is only takes into consideration the direction of changes in the time series. A run is defined as a series of increasing values or a series of decreasing values. The number of increasing or decreasing, values is the length of the run.

$$M = \frac{N(N+1) - \sum_0^2 (n_i)^2}{N} \qquad (5)$$

where M = Expected number of runs,

$n_i$ = Number of price changes of each sign

N = Total number of price changes = $n_0 + n_1 + n_2$

### 4. Data

The sample period chosen for data collection is from September 2005 to June 2015. The data consist of monthly closing values of five leading stock indices of BSE.

### 5. Results and Discussion

**Table 1: Autocorrelation of Monthly Changes in Stock Indices**

| Lag (k) | BSE LargeCap | BSE SmallCap | BSE MidCap | BSE MidSmallCap | BSE LargeMidCap |
|---|---|---|---|---|---|
| 1 | 0.0090 | 0.1257 | 0.1320 | 0.1343 | 0.0362 |
| 2 | -0.0272 | 0.0418 | -0.0101 | 0.0181 | -0.0234 |
| 3 | 0.1041 | 0.0389 | -0.0017 | 0.0208 | 0.0787 |
| 4 | 0.0845 | 0.1044 | 0.0623 | 0.0887 | 0.0825 |
| 5 | 0.0010 | 0.0636 | 0.1308 | 0.0933 | 0.0349 |
| 6 | -0.0638 | -0.0708 | -0.0762 | -0.0745 | -0.0775 |
| 7 | 0.0149 | 0.0336 | 0.0496 | 0.0455 | 0.0238 |
| 8 | -0.0772 | 0.0449 | 0.0184 | 0.0308 | -0.0538 |
| 9 | -0.0046 | -0.0088 | -0.0500 | -0.0242 | -0.0154 |
| 10 | -0.0274 | -0.1210 | -0.0673 | -0.1016 | -0.0349 |
| 11 | -0.1308 | -0.1409 | -0.1314 | -0.1408 | -0.1333 |
| 12 | -0.1058 | -0.0952 | -0.1188 | -0.1020 | -0.1121 |
| 13 | -0.0461 | -0.0688 | -0.0511 | -0.0577 | -0.0421 |
| 14 | -0.2006 | -0.2174 | -0.1771 | -0.2007 | -0.1991 |
| 15 | 0.0452 | -0.0639 | -0.0187 | -0.0441 | 0.0397 |
| 16 | -0.0116 | -0.1345 | -0.1246 | -0.1299 | -0.0322 |
| 17 | -0.1478 | -0.0953 | -0.0997 | -0.0986 | -0.1421 |
| 18 | -0.0271 | -0.1127 | -0.0627 | -0.0893 | -0.0359 |
| 19 | 0.1629 | 0.1040 | 0.0724 | 0.0927 | 0.1458 |
| 20 | -0.0005 | 0.0638 | 0.0498 | 0.0595 | 0.0162 |
| Standard Deviation | 0.08636 | 0.09833 | 0.08758 | 0.09342 | 0.08414 |
| Standard Error | 0.01931 | 0.02199 | 0.01958 | 0.02089 | 0.01881 |

It is observed that for BSE SmallCap, BSE MidCap and BSE MidSmallCap, the autocorrelation coefficient for lag 1 is much larger than twice the standard error. Thus, the autocorrelation differs significantly from zero. From table 1, it is evident that out of the 100 autocorrelations computed for the five stock indices, 70 differ significantly from zero. Thus, 70% of the autocorrelations differ significantly.

The t-values of the autocorrelations corresponding to the five stock indices were computed. It was observed that the autocorrelations are significantly different from zero at 5% level of significance and the corresponding t-values are greater than 1.96. Hence, the stock indices are biased random time series and the stock market is not weakly efficient in pricing securities.

**Table 2: Runs Analysis of Month End Changes in Stock Indices relative to the mean**

| Index | N | $n_0$ | $n_1$ | nruns | Z | p-value |
|---|---|---|---|---|---|---|
| BSE LargeCap | 118 | 59 | 59 | 67 | 1.2019 | 0.2290 |
| BSE SmallCap | 118 | 53 | 65 | 51 | -1.4743 | 0.1403 |
| BSE MidCap | 118 | 52 | 66 | 49 | -1.8137 | 0.0696 |
| BSE MidSmallCap | 118 | 53 | 65 | 49 | -1.8480 | 0.0643 |
| BSE LargeMidCap | 118 | 57 | 61 | 61 | 0.1051 | 0.9159 |

To corroborate this fact, the runs test was performed in the next section. It will test the null hypothesis that price changes are independent. In table 2, N = Total number of observations; $n_0$ = Number of Downs; $n_1$ = Number of Ups; nruns = Total number of observed runs; Z = Standardized Variable test statistic. The results presented in table 2, point to the fact that all the stock indices of BSE show weak form of market efficiency. The information pertaining to yesterday's indices is readily and effectively absorbed by today's indices. Thus, the stocks in the respective indices absorb price information effectively. All the Z values are insignificant at the 5% level.

**Table 3: ADF Test of Month End Changes in Stock Indices**

|  | BSE LargeCap | BSE SmallCap | BSE MidCap | BSE MidSmallCap | BSE LargeMidCap |
|---|---|---|---|---|---|
| **ADF test statistic** | -7.483 | -6.7419 | -6.9681 | -6.8490 | -7.3145 |
| **p-Value** | 0.001 | 0.001 | 0.001 | 0.001 | 0.001 |
| **Included Observations** | 116 | 115 | 115 | 115 | 115 |
| **Number of lags** | 1 | 1 | 1 | 1 | 1 |

ADF tests test the null hypothesis of a unit root in the series of stock prices. If we fail to reject the null hypothesis, then random walk hypothesis couldn't be rejected. The ADF test results clearly show that the null hypothesis is rejected at the 5% significance level for monthly returns of the all the five stock indices.

**TABLE 4: Tests of Normality for Returns of the BSE stock indices: September2005 to June 2015**

**Table 4(a): Kolmogorov-Smirnov Test**

|  | BSE LargeCap | BSE SmallCap | BSE MidCap | BSE MidSmallCap | BSE LargeMidCap |
|---|---|---|---|---|---|
| **Test statistic** | 0.5847 | 0.5897 | 0.6154 | 0.5897 | 0.6154 |
| **p-value** | 0.0000 | 0.0000 | 0.0000 | 0.0000 | 0.0000 |

**Table 4(b): Jarque -Bera Test**

|  | BSE LargeCap | BSE SmallCap | BSE MidCap | BSE MidSmallCap | BSE LargeMidCap |
|---|---|---|---|---|---|
| **Jarque -Bera** | 7.2904 | 41.3902 | 15.5818 | 29.5143 | 9.9741 |
| **JB p-value** | 0.0298 | < 0.0001 | 0.0060 | 0.0012 | 0.0161 |

It is common for the daily returns to not follow normal distribution but not so common for the monthly returns. Both the Kolmogorov-Smirnov test and Jarque-Bera test reject the null hypothesis at the 5% significance level that the monthly returns followed normal distribution during the period of study.

The new government was seen as industrial friendly which boosted the investors' confidence in the market. The foreign investors' were also happy with the change in government which could be seen in the rise in the market. Also, it has been observed that corruption and trading in stock exchanges go hand in hand. A stock market that is functioning well allows people to invest in stocks and make money. It also allows companies to raise capital. But if a stock market is not well regulated, it could pave way for corruption. There are several forms of corruption which could be present in a stock market. They include insider's trading where people who are at top positions and have access to inside information act on it before the information has been revealed to the public. This could evade an investor's faith in the stock market and also be unfair to certain parties. Fraud, false statements and trading abuses are also rampant in non-regulated stock markets. Trading abuses refers to disruptive form of trading in the stock market with the sole objective of exploiting the price of a stock. This could also be a cause of runs in the data for the stock market. This could be a serious issue and should be examined thoroughly.

## 6. Conclusion

The distribution of returns of indices did not follow normal distribution during the period of study which indicates there was no randomness in the stock market. The underlying assumption that the stock prices are random in nature is basic to the Efficient Market Hypothesis and CAPM. In this paper, the evidence of the inefficient form of the Indian Stock Market is presented. From the autocorrelation analysis and the runs test, it is concluded that the series corresponding to the stock indices in the Indian Stock Market are biased random time series. Further, the autocorrelation analysis and ADF test indicate that the behavior of stock prices is such, that the random walk model cannot be applied in the Indian Stock Market. It is also suggested that corruption in the stock market could be a possible reason for the runs. Also, with the shift in governmental regime, the stock market rose as the election results boosted the confidence of the local and global investors. It was shown that there are undervalued securities in the market and the investors can always make excess returns by correctly choosing them.

## 7. Acknowledgement

This work was partially supported by the grant RFFI #14--06-00326
We would like to thank the editor and the anonymous reviewers who have taken time out of their schedule to read and review our manuscript.

## 8. References

1. Malafeyev, O.A., Neverova, E.G.,Nemnyugin, S.A., Alferov, G.V. Multi-criteria model of laser radiation control, Emission Electronics (ICEE), 2014 2nd International Conference on June 30 2014-July 4 2014, DOI: 10.1109/Emission.2014.6893966, Publisher: IEEE
2. Malafeev, O.A., Nemnyugin, S.A. Generalized dynamic model of a system moving in an external field with stochastic components, Theoretical and Mathematical Physics (Impact Factor: 0.7). 01/1996; 107(3):770-774. DOI: 10.1007/BF02070384

3. Malafeyev, O.A., Nemnyugin, S.A., Alferov, G.V. Charged particles beam focusing with uncontrollable changing parameters, Emission Electronics (ICEE), 2014 2nd International Conference on June 30 2014-July 4 2014, DOI: 10.1109/Emission.2014.6893964, Publisher: IEEE

4. XeniyaGrigorieva, Oleg Malafeev A Competitive Many-period Postman Problem With Varying Parameters, Applied Mathematical Sciences, vol. 8, 2014, no. 146, 7249 – 7258

5. Malafeyev, O.A., Redinskikh, N.D., Alferov, G.V. Electric circuits analogies in economics modeling: Corruption networks, Emission Electronics (ICEE), 2014 2nd International Conference on June 30 2014-July 4 2014, DOI: 10.1109/Emission.2014.6893965, Publisher: IEEE

6. Mikio Ito, Akihiko Noda, Tatsuma Wada (2014), International stock market efficiency: a non-Bayesian time-varying model approach, Applied Economics, 46:23, 2744-2754

7. Jasim Al-Ajmi, J. H. Kim (2012) Are Gulf stock markets efficient? Evidence from new multiple variance ratio tests, Applied Economics, 44:14, 1737-1747

8. V. N. Kolokoltsov. O.A. Malafeyev. Understanding Game Theory, Introduction to the analysis of many agent systems of competition and cooperation. World Scientific 2010 (286 pp).

9. V. N. Kolokoltsov, The mathematical Analysis of the Financial Markets, Probability theory and financial mathematics. A brief introduction (textbook, 200 pp), Moscow 2010.

10. Maria Rosa Borges (2011) Random walk tests for the Lisbon stock market, Applied Economics, 43:5, 631-639

11. Truong Dong Loc , Ger Lanjouw, Robert Lensink (2010) Stock-market efficiency in thin-trading markets: the case of the Vietnamese stock market, Applied Economics, 42:27, 3519-3532

12. Malafeev, O.A. Stationary strategies in differential games, USSR Computational Mathematics and Mathematical Physics Volume 17, Issue 1, 1977, Pages 37–46

13. Malafeev O.A. Equilibrium situations in dynamic games, Cybernetics and System Analysis May–June, 1974, Volume 10, Issue 3, pp 504-513

14. Malafeev O.A The existence of situations of ε-equilibrium in dynamic games with dependent movements, USSR Computational Mathematics and Mathematical Physics, Volume 14, Issue 1, 1974, Pages 88-99

15. Bondarenko, L.A., Zubov, A.V. , Zubova, A.F., Zubov, S.V., Orlov, V.B (2015), Stability of quasilinear dynamic systems with after effect, Biosciences Biotechnology Research Asia, Volume 12, Issue 1, Pages 779-788.

16. Anand Pandey (2003), "Efficiency of Indian Stock Market", Indian Economic Journal, Vol.36, No.4 (April – June), 68-121.

17. Chin Wen Cheog (2008), "A Sectoral Efficiency Analysis of Malaysian Stock Exchange under Structural Bank", American Journal of Applied Science Vol.5, No.10, 1291-1295.

18. Yalwar, Y. B. (1988). Bombay Stock Exchange-Rate of Return and Efficiency. Indian Economics Journal, 35(4), 68-121.